# ON THE THEORETICAL AND EXPERIMENTAL EVIDENCES OF MATERIAL PHASE CAUSALITY

I. G. Koprinkov

Department of Applied Physics, Technical University of Sofia, 1756 Sofia, Bulgaria

According to the standard interpretation of quantum mechanics (QM), only amplitude, but not phase, of the state vector is physical observable [1]. In view of some analytic results and critical inspection of matter-wave experiments [2, 3], this appears not to be the case. Here we present theoretical and experimental evidences showing that the material phase (MP) is causally related with the dynamic of quantum system (QS) and may have observable physical consequences.

The internal state dynamic of a QS involved in physical process has been studied, paying special attention to the behavior of the MP. Phase-sensitive dressed states for two-level QS interacting with electromagnetic field of variable amplitude and phase have been obtained [2]. The MP of the real and virtual components of the dressed states shows traceable physical behavior, causally depending on the initial conditions and the relevant physical processes. In the dressed state picture, the MP and the optical phase, whose physical appearance is indisputable, equivalently participate in the formation of the total phase of the quantum state.

The well known "hydrodynamic" formulation of QM reveals [3] fundamental relationship between the amplitude ($R$) and physical action $S$ (*i.e.*, the MP, $\Phi = -S/\hbar$) by means of the coupled differential equations

$$\partial S/\partial t + (\nabla S)^2 / 2m + V(\vec{r},t) - (\hbar^2/2m)(\Delta R/R) = 0$$

$$\partial (R)^2 / \partial t + \nabla \cdot (R^2 \nabla S / m) = 0$$

Consequently, the amplitude and phase are not independent, but codetermine each other. If $R$ becomes related to some element of the physical reality, $\Phi$ ($S$) cannot be irrelevant to that reality. Such $R \leftrightarrow S(\Phi)$ relationship exists *independently on the particular interpretation of QM*, [3]. The acceptance of the MP causality does not necessarily mean to abandon the *epistemological* meaning of the entire wave function. While the later has not physical meaning, its amplitude $R$ and phase $\Phi$ ($S$) may have observable appearances in the physical processes.

It is well established that *matter-wave interference phenomena* take place in the QM phenomenology. These include interference of intraatomic/molecular and external (de Broglie) wavepackets. It is found experimentally that change of the MP may affect the population of given internal quantum state or the interference picture, *i.e.*, leads to observable physical consequences.

The sensible physical behavior and the experimentally observed manifestations show that *the MP is causally related with the dynamic of the QS*. Being an integral quantity, the physical action/phase accumulates contributions from all interactions and motions (independently on the way of motion) of the QS and could be observed in a suitable (interference) experiment. Thus, it is the action/phase, but not other (known) dynamic quantity, that appears to be the most generally related with the physical motion.